\documentclass[useAMS,usenatbib,usegraphicx]{mn2e}

\def\kms{km s$^{-1}$}
\def\msun{M$_{\sun}$}
\def\rsun{R$_{\sun}$}
\def\aap{A\&A}
\def\apjl{ApJ}
\def\apj{ApJ}
\def\apjs{ApJS}
\def\aj{AJ}
\def\mnras{MNRAS}

\def\pasp{PASP}

\def\WD0931{WD 0931+444}
\newcommand {\lessim} {\ {\raise-.5ex\hbox{$\buildrel<\over\sim$}}\ }

\title[A new 20 min Period Binary WD]
{A New Gravitational Wave Verification Source\thanks{Based on observations
obtained at the Gemini and MMT observatories. Gemini is operated by the
Association of Universities for Research in Astronomy, Inc., under a cooperative agreement 
with the NSF on behalf of the Gemini partnership: the National Science Foundation 
(United States), the National Research Council (Canada), CONICYT (Chile), the Australian 
Research Council (Australia), Minist\'{e}rio da Ci\^{e}ncia, Tecnologia e Inova\c{c}\~{a}o 
(Brazil) and Ministerio de Ciencia, Tecnolog\'{i}a e Innovaci\'{o}n Productiva (Argentina).
The MMT is a joint facility of the Smithsonian Institution and the University of Arizona.}}

\author[M. Kilic et al.]
       {Mukremin Kilic$^{1}$,
       Warren R. Brown$^2$,
       A. Gianninas$^1$,
       J. J. Hermes$^3$,
       \newauthor
       Carlos Allende Prieto$^{4,5}$,
       and S. J. Kenyon$^2$\\
       $^1$Department of Physics and Astronomy, University of Oklahoma, 440 W. Brooks St., Norman, OK, 73019, USA\\
       $^2$Smithsonian Astrophysical Observatory, 60 Garden St, Cambridge, MA 02138, USA\\
       $^3$Department of Physics, University of Warwick, Coventry CV4 7AL, UK\\
       $^4$Instituto de Astrof\'{\i}sica de Canarias, E-38205 La Laguna, Tenerife, Spain\\
       $^5$Departamento de Astrof\'{\i}sica, Universidad de La Laguna, E-38206 La Laguna, Tenerife, Spain
}

\begin{document}

\maketitle

\begin{abstract}

We report the discovery of a detached 20 min orbital period binary white dwarf.
\WD0931 (SDSS J093506.93+441106.9) was previously classified as a WD + M dwarf system based on
its optical spectrum. Our time-resolved optical spectroscopy observations obtained at the 8m
Gemini and 6.5m MMT reveal peak-to-peak radial velocity variations of $\approx$400 \kms\
every 20 min for the WD, but no velocity variations for the M dwarf. In addition,
high-speed photometry from the McDonald 2.1m telescope shows no evidence of variability
nor evidence of a reflection effect. An M dwarf companion is physically too large to fit
into a 20 min orbit. Thus, the orbital motion of the WD is almost certainly due to an invisible WD
companion. The M dwarf must be either an unrelated background object or the tertiary
component of a hiearchical triple system. \WD0931 contains a pair of WDs, a 0.32 \msun\ primary and
a $\geq0.14$ \msun\ secondary, at a separation of $\geq0.19$ \rsun. After J0651+2844, \WD0931 becomes the
second-shortest period detached binary WD currently known. The two WDs will lose angular momentum through
gravitational wave radiation and merge in $\leq9$ Myr. The $\log h \simeq -22$ gravitational wave
strain from \WD0931 is strong enough to make it a verification source for gravitational wave
missions in the milli-Hertz frequency range, e.g. the evolved Laser Interferometer Space Antenna ({\em eLISA}),
bringing the total number of known {\em eLISA} verification sources to nine.

\end{abstract}

\begin{keywords}
        binaries: close ---
        white dwarfs ---
        stars: individual (SDSS J093506.93+441106.9, WD 0931+444) ---
        gravitational waves
\end{keywords}

\section{INTRODUCTION}

Short period binary WDs are expected to dominate the gravitational wave foreground at mHz frequencies. 
\citet{nelemans13} predicts $\sim10^8$ double WDs in the Galaxy, including several thousand sources
that should be individually detected by {\em eLISA} \citep{amaro12}.
However, there are only eight {\em eLISA} verification sources currently known.
All but one of these are AM Canum Venaticorum binaries \citep[e.g.,][]{solheim10} with orbital
periods ranging from 5 to 27 min. The remaining system is the 12-min orbital period detached
binary J0651+2844 \citep{brown11,hermes12}. 

Low-mass ($M<0.45$ \msun) WDs are signposts of short period binary systems
\citep[e.g.,][]{marsh95,napiwotzki07}.
We established a radial velocity program, the Extremely-Low Mass (ELM) Survey \citep{brown10,brown12,brown13,kilic10,kilic11,kilic12},
to identify short period binary WDs that are strong gravitational wave sources and potential progenitors
of Type Ia and .Ia supernovae \citep{bildsten07,shen09,kilic14}. The ELM Survey has so far discovered 55
binaries, all with $P\lessim 1$ d, including 32 that will merge within a Hubble time \citep{gianninas14}. Three of these have orbital periods less than an hour;
J0651+2844, J0106$-$1000, and J1630+4233.
 
We have recently extended our search for ELM WDs to the Sloan Digital Sky Survey Data Release 10
\citep[SDSS DR10,][]{ahn14}. We fit all available DR10 spectroscopy for stellar sources with
WD model spectra \citep{koester10} to identify ELM WDs with $5\leq\log{g}\leq7$ and $M\leq0.3$
\msun. Here we present optical spectroscopy and photometry of one of these sources, the relatively
bright ($g=17.7$ mag) SDSS J093506.93+441106.9 (\WD0931).

\citet{usher82} identified
\WD0931\ as a UV-excess object in the Palomar Schmidt plates, and \citet{mitchell04} confirmed
it as a DA WD based on optical spectroscopy. However, the SDSS observations clearly show a composite spectrum
of a DA WD plus an M dwarf \citep{kleinman13}. \citet{silvestri06}, \citet{mansergas07,mansergas10},
and \citet{heller09} classify \WD0931 as a low-mass WD with $T_{\rm eff}\sim20,000$K, $\log{g}\approx7$, and
an M1 dwarf companion. \citet{mansergas07} find an average mass and radius of $M=0.46$ \msun\ and
$R=0.43 \pm 0.09$ \rsun\ for M1 dwarfs, which implies a distance of 1440 $\pm$ 340 pc for the M dwarf in \WD0931.
\citet{mansergas10} revise the distance to $\approx$1150 $\pm$ 220 pc, which is 2.2$\sigma$ farther
away than the WD (see \S3.1). However, there seems to be a systematic effect that leads to overestimating
the M dwarf distances in DA + M systems \citep[see][]{mansergas10}. 
For example, the distance estimates for the WDs and M dwarfs in the 58 Post-Common-Envelope-Binaries
presented in \citet{nebot11} differ by as much as 3$\sigma$. 
Hence, the inconsistent distance estimates for the WD and M dwarf do not rule out physical association.
\citet{mansergas07} also point out that the M dwarf in \WD0931\ has a poorly defined Na I doublet, possibly
due to orbital motion. If so, the relatively hot WD would cause significant
heating of the secondary star, which would be detected as a reflection effect in the optical lightcurve.

Our radial velocity and high-speed photometry demonstrate that \WD0931\ contains a
pair of WDs with an orbital period of only 20 min, and that the M dwarf is not a member of
this binary. In Section 2 we describe our spectroscopic and photometric observations. 
In Sections 3 and 4 we constrain the physical parameters of this system and discuss
the nature and future evolution of \WD0931. We conclude in Section 5.

\begin{figure}
\vspace*{-0.6in}
\hspace*{-0.2in}
\includegraphics[width=2.6in,angle=-90]{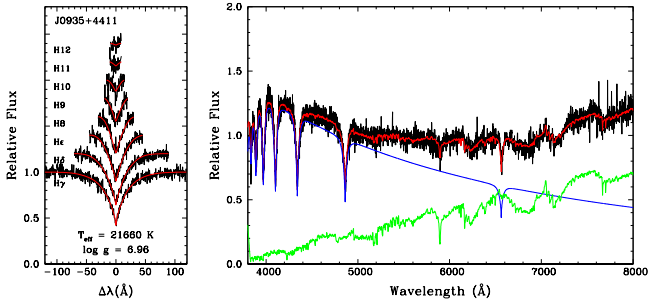}
\vspace*{-0.5in}
\caption{Left Panel: Model fits (red lines) to the Balmer line profiles of \WD0931\ from the MMT spectroscopy.
Right Panel: Model fits to the SDSS spectroscopy, including the contribution from the WD (blue line) and M dwarf (green line).
\label{fig:spec}}
\end{figure}

\section{OBSERVATIONS}

We used the 6.5m MMT with the Blue Channel spectrograph to obtain medium resolution spectroscopy
of \WD0931\ in 2013 November and 2014 March. The former observations used
the 832 line mm$^{-1}$ grating in second order, providing wavelength coverage from 3600 \AA\ to 4500 \AA\
and a spectral resolution of 1.2 \AA. The latter observations used the 1200 line mm$^{-1}$, providing
wavelength coverage from 5640 \AA\ to 6940 \AA\ and a spectral resolution of 1.5 \AA.
We obtained all observations at the parallactic angle.

We obtained follow-up optical spectroscopy using the 8m Gemini-North telescope with
the Gemini Multi-Object Spectrograph (GMOS) in 2014 March as part of program GN-2013B-DD-9.
We obtained a sequence of 32$\times$150s exposures with the R831 grating and a 0.5$\arcsec$ slit, providing
wavelength coverage from 5460 \AA\ to 7560 \AA\ and a resolving power of 3720.
We also obtained a sequence of 20$\times$120 s exposures with the B1200 grating and a 0.5$\arcsec$ slit,
providing wavelength coverage from 3700 \AA\ to 5150 \AA\ and a resolving power of 3635.
Each spectrum has a comparison lamp exposure taken within 10 min of the observation time. 
We flux-calibrate using blue spectrophotometric standards \citep{massey88}, and measure radial velocities
using the cross-correlation package RVSAO. 

We acquired high speed photometry of \WD0931\ using the McDonald 2.1m Telescope with the
Puoko-nui North camera \citep{chote14} over five nights in 2013 December. We
obtained images through both a BG40 filter (8.2 h) and an SDSS $z-$band filter (5.8 h) to look for
variations in the light curve for the WD and M dwarf, respectively.

\vspace*{-0.2in}
\section{RESULTS}

\subsection{The ELM WD}

Figure \ref{fig:spec} shows the optical spectrum of \WD0931\ along with our model fits.
The optical spectrum is dominated by the WD below 5000 \AA, which enables us to use
the Balmer lines to constrain the physical parameters of the WD precisely.
We use the H$\gamma$-H12 lines in our MMT spectra
and an extended model atmosphere grid based on the \citet{bergeron95} models,
with recent improvements presented in \citet{tremblay09},
to constrain the atmospheric parameters of the WD.
The best-fit model has $T_{\rm eff} = 21660 \pm 380$ K and $\log{g} = 6.96 \pm 0.05$.
The recent evolutionary calculations by \citet{althaus13} indicate that
\WD0931\ is an $M=0.32 \pm 0.02$ \msun\ and $R=0.031 \pm 0.003$ \rsun\ WD at a distance of 660 $\pm$ 70 pc.

\begin{table}
\centering
\caption{Radial velocity measurements for \WD0931. The full table is available online.}
\begin{tabular}{cr}
\hline
HJD$-$2456500 & $v_{helio}$ \\
(days) & (\kms) \\
\hline
97.029593  &   104.34 $\pm$  23.31 \\
98.029345  &   191.50 $\pm$  30.83 \\
99.023102  &   199.16 $\pm$  25.51 \\
99.023970  &    84.95 $\pm$  35.74 \\
99.024850  & $-$16.74 $\pm$  39.36 \\
\hline
\end{tabular}
\end{table}

The right panel in Figure \ref{fig:spec} shows composite WD + M dwarf model fits to the SDSS spectrum
of \WD0931. We use the best-fit WD model to subtract out
the contribution from the WD and use the \citet{bochanski07} templates to fit the contribution
from the M dwarf. The best-fit template is M1.5, consistent with the previous analyses 
by \citet{silvestri06}, \citet{heller09}, and \citet{mansergas10}. This spectral type
is also consistent with the infrared data from the Two Micron All Sky Survey \citep{cutri03} and the Wide-Field
Infrared Survey Explorer \citep{wright10}.

\vspace*{-0.2in}
\subsection{The Orbital Period}

\begin{figure}
\vspace*{-0.15in}
\hspace{-0.4in}
\includegraphics[width=2.0in]{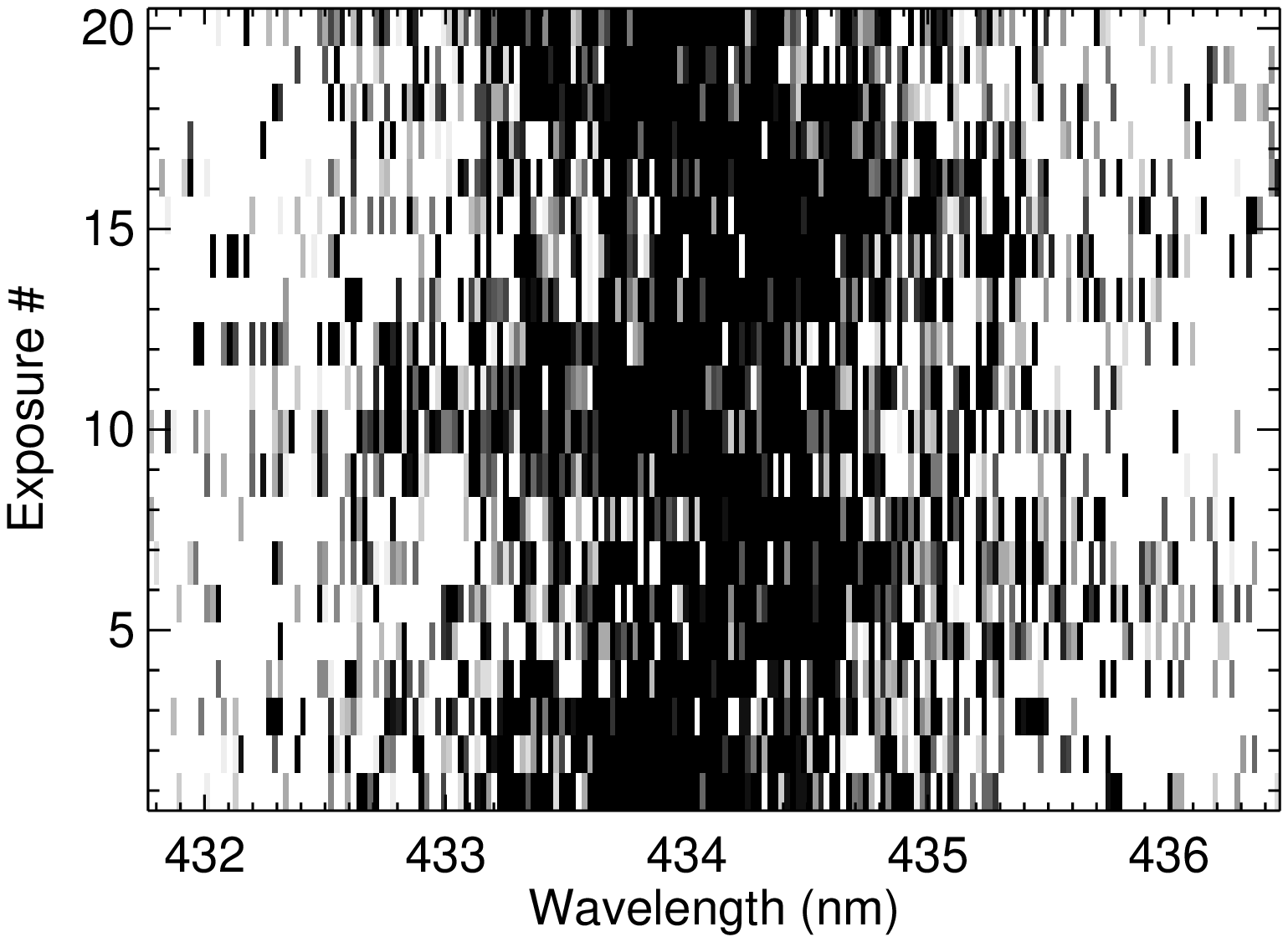}\hspace*{-0.2in}
\includegraphics[width=2.0in]{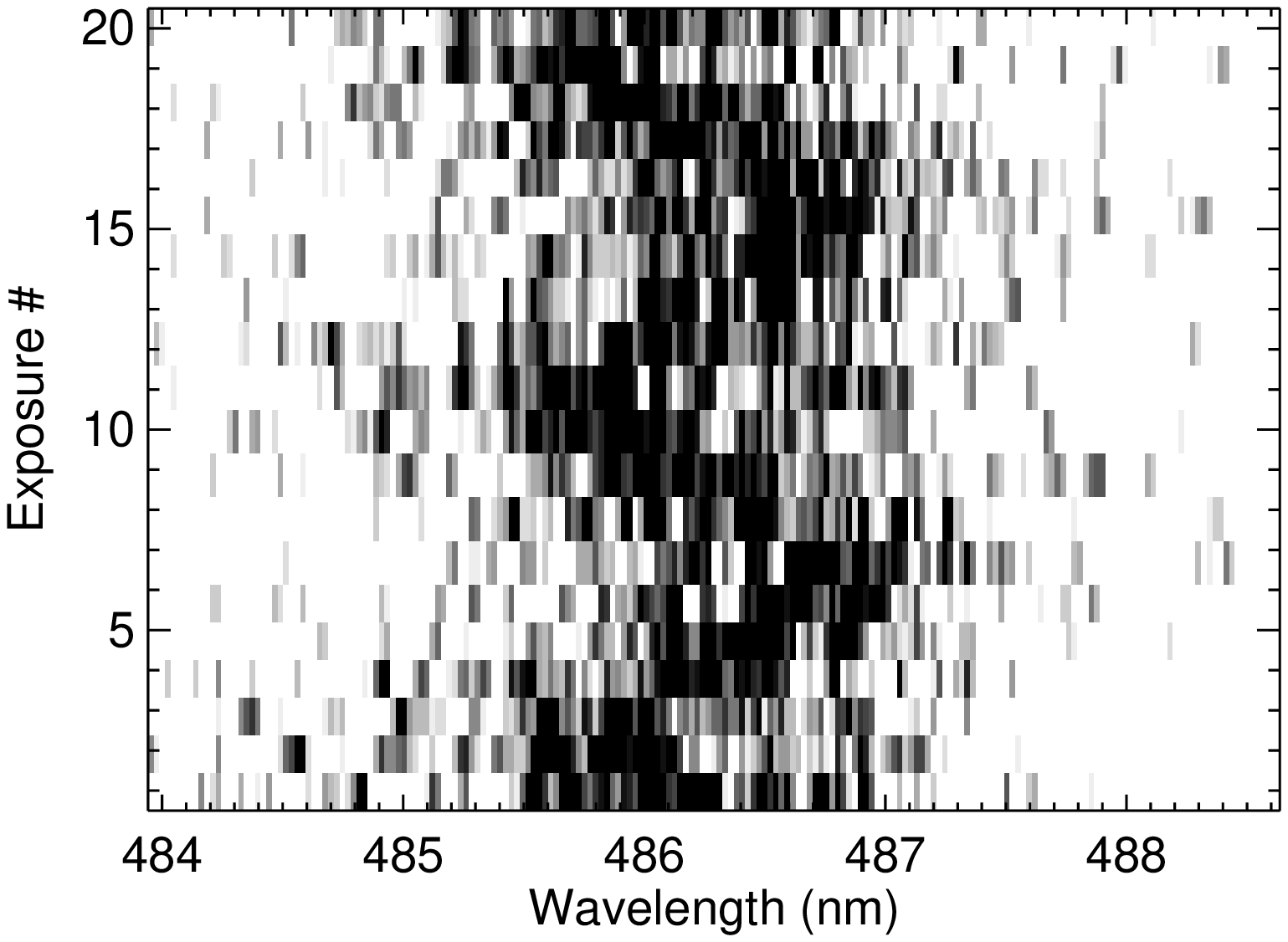}
\caption{Gemini time-resolved spectroscopy of H$\gamma$ (left panel) and H$\beta$
lines (right panel) over 45 min. Both lines clearly show a 20 min periodicity.
\label{fig:blue}}
\end{figure}

Figure \ref{fig:blue} shows Gemini time-resolved spectroscopy of H$\gamma$ and H$\beta$ lines
over 45 min. H$\beta$ and higher order Balmer lines are relatively clean, with no evidence
of significant contamination or activity from the M dwarf. All of the Balmer lines clearly show evidence
of a 20 min orbital period. 

\begin{figure}
\vspace*{-0.2in}
\hspace*{0.2in}
\includegraphics[width=3.0in,angle=0]{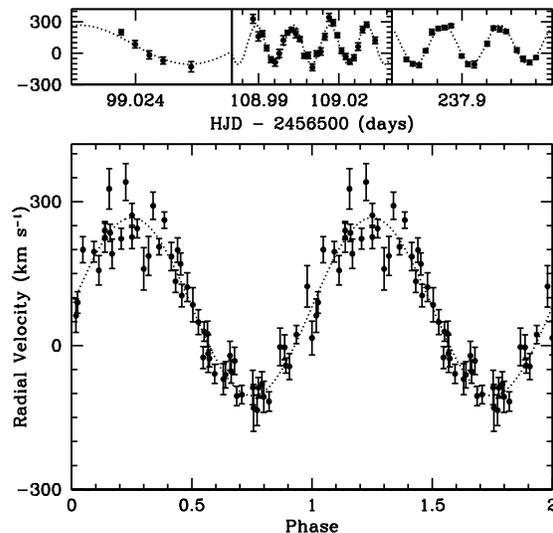}
\vspace*{-0.2in}
\caption{The radial velocities of the Balmer lines in \WD0931.
The bottom panel shows all of these data points phased with the best-fit
period. The dotted line represents the best-fit model for a circular orbit with a period
of 0.01375 d.
\label{fig:rv}}
\end{figure}

Figure \ref{fig:rv} shows 55 radial velocity measurements for \WD0931\ obtained from
H$\gamma$ through H12 lines. Table 1 lists these velocities.
We compute the best-fit orbital elements using the code of \citet{kenyon86}, and
perform a Monte Carlo analysis to verify the uncertainties in the orbital parameters
\citep[see][for details]{brown12}.
\WD0931 exhibits radial velocity variations with a semi-amplitude of $K = 185.4 \pm 3.2$
\kms\ and orbital period of $P = 0.01375 \pm 0.00051$ d, or 19.8 $\pm$ 0.7 min, with a significant
alias at 20.4 min.
The observed velocity amplitude is underestimated, however, because our 2 min long exposures span 10\%
of the orbital phase. The corrected velocity semi-amplitude is $K=198.5 \pm 3.2$ \kms.
After correcting for the gravitational redshift of the WD (6.6 \kms), the systemic velocity
is 74.3 $\pm$ 2.3 \kms. The mass function is $f= 0.0111 \pm 0.0007$ \msun, which
implies an $M\geq$0.13 \msun\ companion at a separation of 0.19 \rsun.

At an inclination angle of $i=90^{\circ}$, a 0.13 \msun\ companion would have a Roche-lobe radius
of 0.056 \rsun\ \citep{eggleton83}. This is smaller than the radii for all known M dwarfs
\citep[see Table 5 in][]{mansergas07}. Hence, if the orbital motion of \WD0931\ is due to
an M dwarf, such a companion would fill its Roche lobe, yet
there is no evidence of mass transfer in this system. In fact, no main-sequence star can fit into
this orbit; the orbital period of this system is significantly shorter than the period minimum
for cataclysmic variables \citep[$\approx78$ min,][]{hellier01}. Clearly, the visible M dwarf
in \WD0931\ cannot be a binary companion of the WD. Based on the mass function alone, the
probability of a $M\leq1.4$ \msun\ companion is 97.5\%. Hence, the companion is almost
certainly another WD.

\vspace*{-0.2in}
\subsection{The M dwarf in \WD0931}

\begin{figure}
\vspace*{-0.17in}
\hspace{-0.3in}
\includegraphics[width=2.0in]{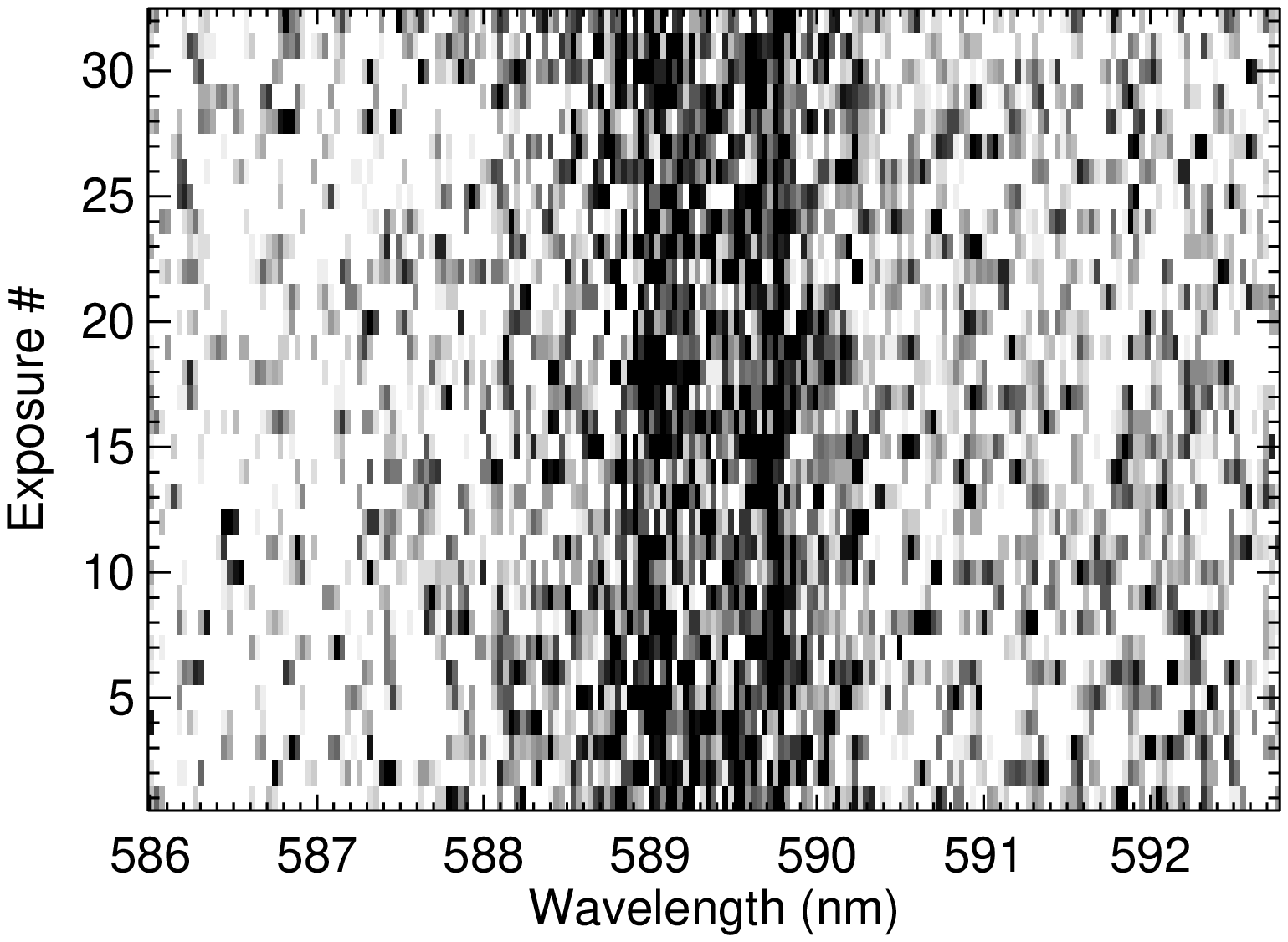}\hspace*{-0.2in}
\includegraphics[width=2.0in]{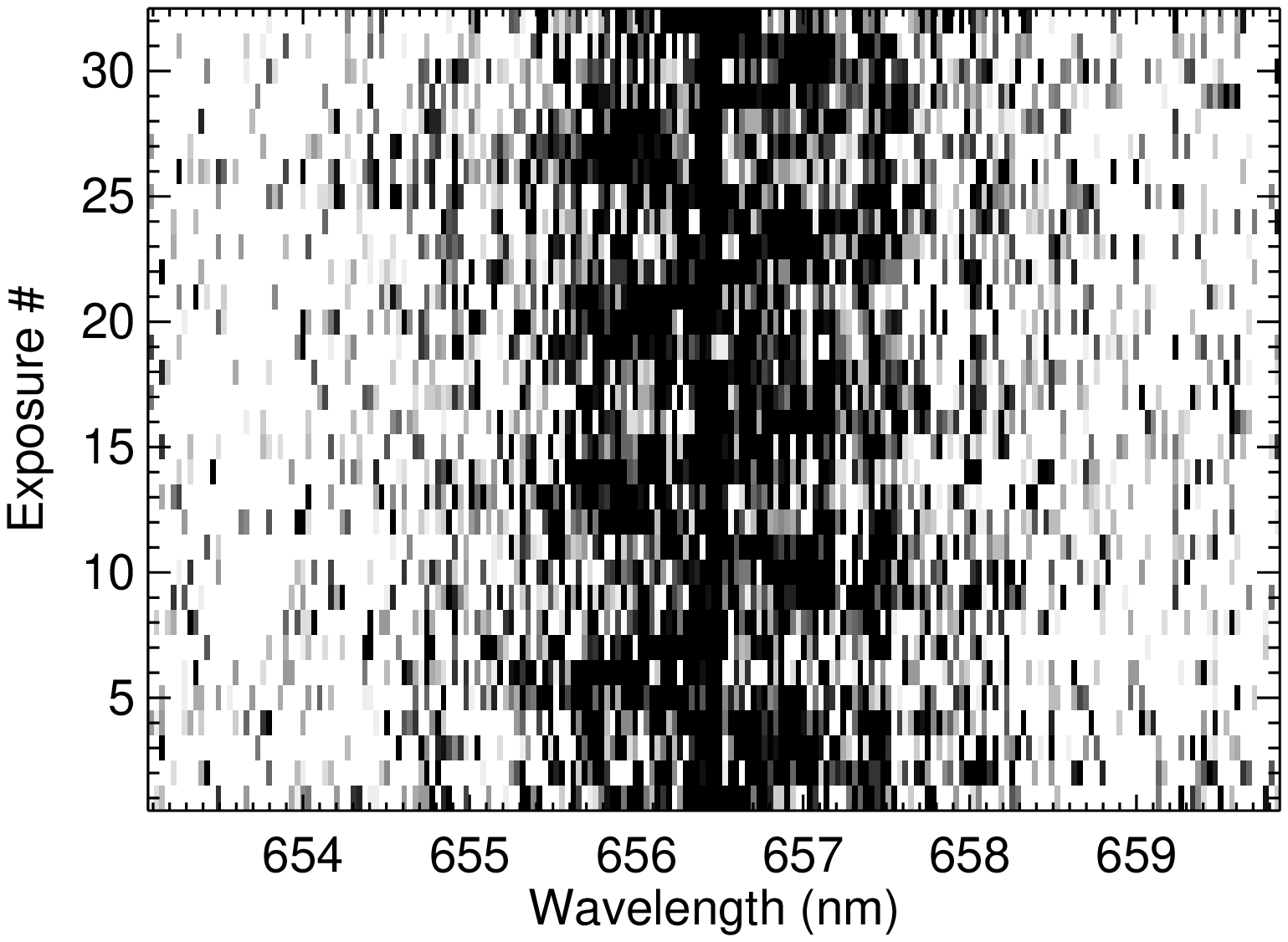}
\caption{Gemini time-resolved spectroscopy of the Na I doublet (left panel) and the H$\alpha$
line (right panel) over 90 min. The Na I lines and the H$\alpha$ line from the M dwarf are stationary, whereas
the H$\alpha$ line from the WD clearly shows a 20 min periodicity.
\label{fig:sodium}}
\end{figure}

Figure \ref{fig:sodium} shows the Gemini time-resolved spectroscopy of the Na I doublet and
the H$\alpha$ line. There is no evidence of any metal-pollution in the DA WD,
hence
the Na I doublet is clearly from the M dwarf. If the M dwarf was associated with the 20 min binary, given
the mass ratio of the two stars, we would expect to see peak-to-peak velocity variations of
276 \kms. On the contrary, the velocity of the Na I doublet is consistent with $\Delta v=0 \pm 4$
\kms\ in all of the Gemini and MMT spectra. Forcing a 20-min periodicity, the best-fit velocity
semi-amplitude is 3.3 \kms. Clearly, the M dwarf does not display any significant radial velocity variability. 

The right panel of Figure \ref{fig:sodium} is perhaps more revealing. This figure displays two components
for the H$\alpha$ line. The first one, from the M dwarf, is stationary.
While the second one, from the WD, shows significant velocity variations.
The H$\alpha$ line from the WD clearly goes through 4+ orbital cycles in this figure,
consistent with the orbital period of 20 min, as measured from the rest of the Balmer lines (see Fig.
\ref{fig:blue}).

\vspace*{-0.2in}
\subsection{The Light Curve}

\begin{figure}
\includegraphics[width=3.0in]{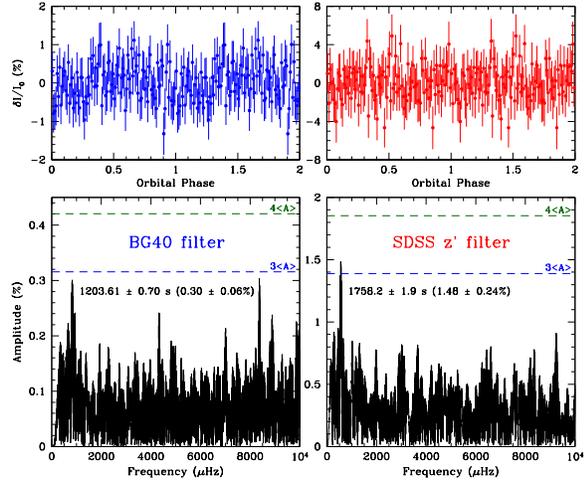}
\caption{High speed photometry of \WD0931\ in the BG40 (left panel) and the $z-$band filter (right panel).
The light curve is binned into 100 phase bins and folded at the best-fit orbital period from the radial velocity
measurements. The bottom panels show the fourier transform of both datasets.
\label{fig:lc}}
\vspace*{-0.1in}
\end{figure}

Figure \ref{fig:lc} shows the McDonald 2.1m telescope light curve of \WD0931. The fourier transforms of the
blue-sensitive BG40 and the red-sensitive SDSS $z-$
band data are also shown. These light curves do not reveal any significant photometric
variability. There is a marginal peak
($<3\sigma$) at 1203.6 s in the BG40 filter data. This is consistent with variations at the orbital
period, which may be due to the relativistic beaming effect. Similarly there is a marginal
peak in the $z-$band data at 1758.2 s. No variations are expected at that frequency, hence this marginal
signal is most likely due to noise from atmospheric variability in the $z-$band.

Assuming that the secondary star has a radius comparable to the ELM WD, the lack of eclipses in our
photometry require $i\leq 70^{\circ}$, which implies a $M\geq0.14$\msun\
WD companion. For $i\leq 70^{\circ}$ and the limb- and gravity-darkening coefficients of $u=0.34$
\citep{gianninas13} and $\tau=0.48$, respectively, we expect $\leq0.25$\% ellipsoidal
variations in the BG40 filter due to tidal distortions of the
primary. This amplitude is below our detection threshold, so the lack of a signal at half the orbital
period does not provide any new constraints on the inclination of the binary.

The absence of a reflection effect is additional evidence that the M dwarf does not orbit the WD.
For comparison, HS 2043+0615 is a $T_{\rm eff}\approx26,000$ K subdwarf star with a
0.18-0.34 \msun\ M dwarf companion in a 0.3 d binary. This system shows 0.15 mag brightness variations
due to the reflection effect \citep{geier14}. Similarly, SDSS J162256.66+473051.1 contains a subdwarf star
with a heated brown dwarf companion in a 0.07 d orbit, and shows a $\sim$10\% reflection effect
\citep{schaffenroth14}. Yet, there is no evidence of a reflection effect for \WD0931\ in our
photometry. This further reinforces our determination that the M dwarf is not in the 20-min orbit with
the visible WD.

\vspace*{-0.2in}
\section{DISCUSSION}

\subsection{A New Verification Binary}

Our optical spectroscopy and photometry demonstrate that \WD0931\ is a 20 min orbital
period detached double WD. Along with J0106$-$1000, J0651+2844, and J1630+4233,
\WD0931\ becomes the fourth detached WD binary known to have a period less than
an hour. The two WDs in this system will merge in $\leq$9 Myr. After J0651,
\WD0931\ becomes the second quickest WD merger system currently known. 

At a distance of 660 pc and $i\leq 70^{\circ}$, we expect the gravitational wave strain at Earth
$\log h \geq -22.17$ at $\log \nu$ (Hz) = $-2.77$ \citep{roelofs07}. Hence, \WD0931\ is clearly a
verification source for {\em eLISA}, bringing the total number of known {\em eLISA} verification
sources to nine.

When the mass transfer starts, \WD0931 will most likely have unstable mass transfer \citep{marsh04}.
However, the merger outcome is uncertain because of the unknown inclination angle and the companion mass.
For a relatively high inclination angle of $i\approx70^{\circ}$, the companion would be a very
low-mass WD with $M=0.14$\msun, and the merger will lead to a single He-burning subdwarf star. 
For an inclination of $i=31^{\circ}$, \WD0931\ would be an equal mass binary with a merger
time of 4.4 Myr, and $\log h = -21.6$. This is similar to the gravitational wave strain
from the verification binary HP Lib \citep{nelemans06}. The non-detection of the secondary WD
in our spectroscopy implies that the companion is cooler \citep[like CS 41177,][]{bours14}
and/or more massive \citep[like J0651,][]{brown11}. Therefore, the gravitational wave strain of
\WD0931\ may be even higher than this estimate, potentially making it one of the best verification
sources known. 

\vspace*{-0.2in}
\subsection{Is \WD0931 a Triple System?}

There is growing interest in triples containing an inner double WD binary. 
\citet{perets12} suggest that mass loss in an evolving triple system leads to orbital instability,
triggering close encounters or collisions in the system. This mechanism could explain prompt
supernovae Ia, and contribute to the Ia event rate \citep{kushnir13}. However, \citet{hamers13}
find this rate to be low.

To explore the eccentricity of the \WD0931\ binary, we fit both circular and eccentric orbits to
the radial velocity data. The 3$\sigma$ upper limit on eccentricity is $e=0.02$, and the
\citet{lucy71} test strongly prefers a circular orbit. This suggests that the Kozai mechanism
is not important for \WD0931, and we are yet to find a triple system containing an eccentric
double WD inner binary.

Given the size of the orbit (0.19\rsun), and the lack of evidence for mass transfer and a
reflection effect, the M dwarf that contaminates the SDSS spectrum of \WD0931\ cannot be a
member of the 20 min period binary. The distance estimate also puts the M dwarf significantly
further away than the WD, though the distance uncertainty is rather large. 
Cross-correlating with a template spectrum of an inactive M1 dwarf \citep{bochanski07},
we measure velocities of $-47 \pm 9$ and $-37 \pm 13$ \kms\ for the Na I doublet from the Gemini and
MMT spectra, respectively. These are significantly different than the systemic velocity of
the WD binary (74.3 $\pm$ 2.3 \kms). Hence, the M dwarf is likely a background source.

Based on six epochs from the USNO-B and the SDSS, \citet{munn04} measure a proper motion of
($\mu_{\alpha} cos \delta, \mu_{\delta}) = (-19.9,-0.8$) mas yr$^{-1}$.
\WD0931\ is unresolved in the Palomar plates and the SDSS images. If the proper motion measurement
is correct, \WD0931\ would have moved only 1$\arcsec$ between the first Palomar observations and the SDSS.
Hence, the current proper motion measurements are inconclusive in deciding if the M dwarf is the tertiary
component of the \WD0931 system. If the M dwarf is a background object, the Hubble Space Telescope observations
may be able to resolve the WD and the M dwarf and confirm or rule out any physical association based on
common proper motion.

\vspace*{-0.15in}
\section{CONCLUSIONS}

We identify \WD0931\ as a new {\em eLISA} verification source, only the ninth such system known.
\WD0931\ is a 20 min orbital period detached binary WD, the second-quickest merger system currently
known. It contains a 0.32 \msun\ WD and a $\geq0.14$ \msun\ companion. The M dwarf in this
system may or may not be associated with \WD0931, and we propose follow-up proper motion measurements
to distinguish between the two scenarios. Since the companion WD is not detected in our spectroscopy
or photometry, it is likely cooler and/or more massive than the ELM WD. A more massive companion would
make it one of the strongest sources of gravitational radiation currently known. 

\vspace*{-0.2in}
\section*{Acknowledgements}

We gratefully acknowledge the support of the NSF under grant AST-1312678.
JJH acknowledges funding from the European Research Council under the European Union's
Seventh Framework Programme (FP/2007-2013) / ERC Grant Agreement n. 320964 (WDTracer).
We thank the Gemini staff, including Atsuko Nitta and Nancy Levenson, for
help with our observing program.

\end{document}